\documentclass[twocolumn,aps,prl,showpacs,groupedaddress,amsfonts,amssymb,amsmath]{revtex4-1}
\usepackage{graphicx}
\usepackage{epstopdf}
\usepackage{array}
\usepackage[colorlinks=true, letterpaper=true, pdfstartview=FitV, linkcolor=blue, citecolor=blue, urlcolor=blue]{hyperref}
\usepackage{color}

\begin{document}

\draft

\title{Electric-field control of pure spin photocurrent in germanene}
\author{Yaqing Yang,$^{1}$ Zhen Zhang,$^{2}$ Liwen Zhang,$^{3}$ Liantuan Xiao,$^{1,4}$ Suotang Jia,$^{1,4}$ Jun Chen,$^{2,4,*}$ and Lei Zhang$^{1,4,\dagger}$ }
\address{$^1$State Key Laboratory of Quantum Optics Technologies and Devices, Institute of Laser Spectroscopy, Shanxi University, Taiyuan 030006 China\\
$^2$State Key Laboratory of Quantum Optics and Quantum Optics Devices, Institute of Theoretical Physics, Shanxi University, Taiyuan 030006, China\\
$^3$School of Physics and Information Engineering, Shanxi Normal University, Taiyuan 030031, China\\
$^4$Collaborative Innovation Center of Extreme Optics, Shanxi University, Taiyuan 030006, China}

\begin{abstract}
The electrical control of pure spin current remains a central challenge in spintronics, particularly in time-reversal symmetric systems composed of nonmagnetic elements, where spin and electric fields interact only indirectly. In this work, we develop a theoretical framework for electrically tuning pure spin photocurrent in two-dimensional materials with time-reversal symmetry via a gate electric field. Through theoretical analysis, we demonstrate that in systems with spin-orbit coupling and in-plane mirror symmetry, an out-of-plane electric field induces spin splitting and reversal in the band structure near the Fermi energy, enabling magnitude control and direction reversal of the pure spin photocurrent. To validate this mechanism, we perform first-principles calculations on germanene, an experimentally realized two-dimensional material. Beyond amplitude modulation, we reveal that reversing the direction of the applied electric field leads to a corresponding reversal of the pure spin photocurrent. Furthermore, we show that the pure spin photocurrent can be tuned by varying the photon energy and the incident angle of light, providing additional degrees of control over spin transport. These findings establish a robust strategy for electric-field-controlled pure spin transport in two-dimensional materials, offering new possibilities for the development of optospintronic devices.

\end{abstract}

\maketitle
\emph{Introduction}-With the continuous advancement of information technology, the demand for low-power, high-speed, and highly integrated electronic devices has grown significantly.\cite{zwyssig2008,ren2018,nazir2020} In this context, spintronics has emerged as a compelling alternative to traditional charge-based electronics, offering a pathway toward energy-efficient information processing by utilizing the electron's spin degree of freedom for data storage and transport.\cite{gardelis1999,wolf2001,vzutic2004,candini2011,hirohata2014,otani2017,ahn2020,stevens2003,hoffmann2007,yang2008,tang2013,tao2020,puebla2020,hirohata2020,yao2025} Despite substantial progress, key challenges remain unresolved, for instance, the efficient generation and tunable control of pure spin currents that carry spin angular momentum without net charge flow. Although mechanisms such as the spin Hall effect, spin Seebeck effect, and spin pumping can generate pure spin currents, their practical utility is often constrained by low efficiency, material limitations, and limited tunability.\cite{dyakonov1971,sinova2015,jung2012,saitoh2006,uchida2008,adachi2013,qu2013,matsuo2018,du2014,zhang2012,shiomi2014,rimmler2025,bai2024,zheng2023,dal2024} Another central challenge is the precise control of spin direction, particularly for pure spin current. Traditional approaches employ external magnetic fields, which couple directly to spin angular momentum via the Zeeman interaction.\cite{wolf2001,chernyshov2009,tanttu2019,tang2023} However, as device dimensions shrink and energy efficiency becomes increasingly critical, magnetic-field-based techniques face scalability and integration challenges. An alternative strategy leverages spin-orbit coupling (SOC), which intrinsically links an electron’s spin to its momentum, allowing electric fields to act as effective magnetic fields and thus enabling all-electrical spin control without magnetic components.\cite{Zumb2002,manchon2015,Flindt2006,galitski2013} In parallel, optical approaches have gained traction as versatile tools in spintronics, offering advantages including contactless operation, ultrafast response, high spatial resolution, and compatibility with a wide range of materials.\cite{tarasenko2005,tao2020,zhang2021,xu2021,lihm2022} Particularly, in materials with specific crystalline symmetries, pure spin photocurrents can be generated via nonlinear optical processes, especially along mirror-symmetric directions or in systems exhibiting combined parity-time (PT) symmetry.\cite{tao2020,shan2015,zhao2022,rahimi2022,mu2021}

Despite significant progress in generating pure spin photocurrent, for instance, tuning the direction of third-order pure spin currents with dual-color light and achieving effective control of their magnitude,\cite{bhalla2024} the tuning of photovoltaic effect generated pure spin current directions—particularly through electric fields—remains largely unexplored. Existing efforts on electric-field control of pure spin photocurrent have predominantly focused on systems with broken time-reversal symmetry, such as antiferromagnetic graphene nanoribbons.\cite{zhang2021,Jiang2021} In contrast, achieving effective modulation and reversal of pure spin photocurrent in time-reversal symmetric, nonmagnetic materials remains a fundamental challenge. This gap raises a key question central to the development of spin-optoelectronic technologies: Can electric fields be harnessed to control both the magnitude and direction of pure spin photocurrent in nonmagnetic systems that preserve time-reversal symmetry?

\begin{figure*}
	\centering
	\includegraphics[scale=0.63]{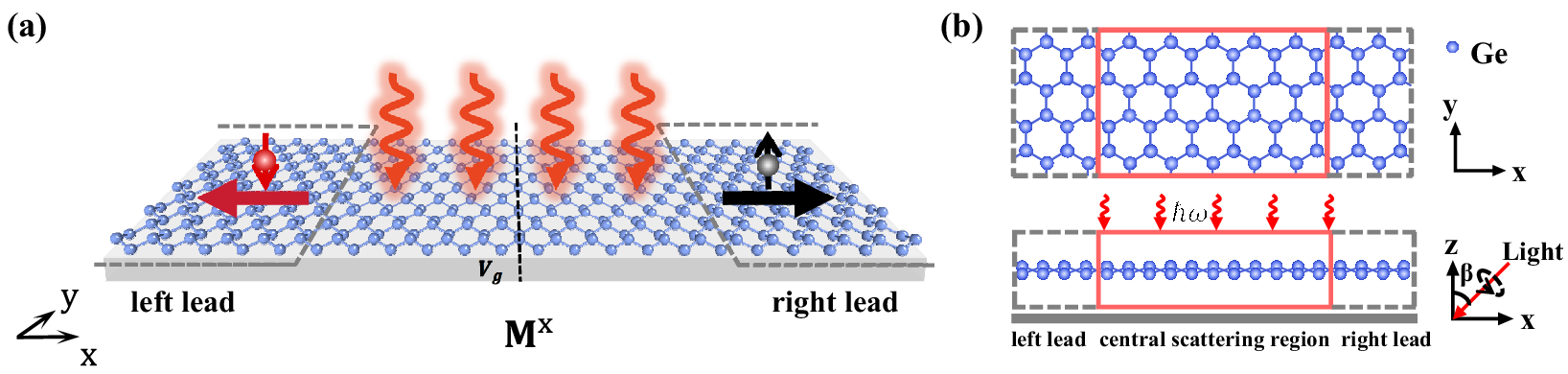}
	\caption{(a) Schematic illustration of electric-field-tuned pure spin photocurrent generation via nonlinear optical effects in two-dimensional systems with mirror symmetry $M^x$. (b) Schematic representation of a two-probe device based on germanene. The system comprises three regions: left and right leads, and a central scattering region where incident light with photon energy $E=\hbar\omega$ is applied.}
	\label{Fig1}
\end{figure*}

In this work, we answer this question affirmatively by developing a theoretical framework for electrically tuning pure spin currents in time-reversal symmetric, nonmagnetic systems. Using theoretical analysis of a Dirac Hamiltonian and the formalism of nonlinear response theory, we show that in mirror-symmetric systems, an out-of-plane electric field can effectively modulate pure spin photocurrent via the photovoltaic effect. This modulation requires the presence of in-plane mirror symmetry and an electric-field-induced spin polarization near the Fermi level. We demonstrate that the applied electric field not only tunes the magnitude of the spin photocurrent but also reverses its direction. To validate our predictions, we perform first-principles calculations based on the non-equilibrium Green’s function (NEGF) method combined with density functional theory (DFT), and confirm this mechanism in germanene, a two-dimensional material that has been experimentally realized. Under linearly polarized light, the pure spin photocurrent in germanene exhibits robust tunability with respect to both the electric field and optical parameters. In particular, we find that the photocurrent response can be further modulated by varying the photon energy and incident angle of the light. These results establish a viable route for the electric-field control of pure spin photocurrent in two-dimensional nonmagnetic systems, offering new opportunities for the design of volatile spintronic memory and optospintronic devices.

\emph{Theoretical formalism}-We begin by identifying the necessary conditions for the generation of a pure spin photocurrent along the $a$ direction, which are:
\begin{equation}
	\begin{aligned}
		\label{J_s}
		J^{a,s^z}\neq0,
		J^{a,s^0}=0.
	\end{aligned}
\end{equation}
Here, $J^{a,s^z}$ denotes the spin photocurrent, while $J^{a,s^0}$ corresponds to the charge photocurrent. The spin and charge nonlinear optical (NLO) photocurrent induced by linearly polarized light of frequency $\omega$ are given by:
\begin{equation}
\label{NLO}
J^{a,s^i}=\sum_{\Omega=\pm \omega}\sigma^{a,s^i}_{bb}(0;\Omega,-\Omega)E^b(\Omega)E^b(-\Omega).
\end{equation}
where, $i$ equals $z$ or $0$ to distinguish the spin and charge components, respectively, and $b$ denotes the polarization direction of the optical electric field. The quantity $E^{b}(\omega)$ is the Fourier component of the optical electric field at frequency $\omega$ along the $b$ direction. $\sigma^{a,s^i}_{bb}$ represents the NLO conductivity tensor, which can be expressed as:\cite{kraut1979,von1981,sipe2000,Young2013,Parker2019,xu2021,Shi2023}
\begin{equation}
	\begin{aligned}
&\sigma^{a,s^i}_{bb}(0;\omega,-\omega)=-\frac{e^3}{\hbar^2 \omega^2} \int\\
& {\frac{d\bf k}{(2\pi)^3}} \sum_{m,n,l} \frac{f_{lm} v^b_{lm}}{\omega_{ml}-\omega+\frac{i}{\tau}} \left(\frac{j^{a,s^i}_{m n} v^b_{nl}}{\omega_{mn}+\frac{i}{\tau}}-\frac{j^{b,s^i}_{mn} v^a_{nl}}{\omega_{nl}+\frac{i}{\tau}}\right),\label{sigma}
    \end{aligned}
\end{equation}
Here, $f_{lm}=f_l-f_m$ reprsesents the Fermi-Dirac occupation difference, and $\omega_{lm}=\omega_l-\omega_m$ is the energy difference between bands $l$ and $m$, while $\tau$ represents the carrier lifetime. Note that we include only intraband carrier lifetimes and assume a uniform lifetime for all carriers. For treatments incorporating interband contributions to the lifetime, we refer the reader to Ref. \cite{Bhalla2023}. The velocity matrix element along the $a$ direction is given by $v^a_{lm}=\langle l|{\partial_{k_a} H} |m\rangle$, where $H$ is the system Hamiltonian.\cite{xu2021} The spin current operator $j^{a,s^i}$ for $i=z$ is given by $j^{a,s^z}=\frac{1}{2}(v^as^z+s^zv^a)$, where $s^z=\frac{\hbar}{2}\sigma^z$ is the spin operator. For $s^0=e$, $j^{a,s^0}$ corresponds to charge current. When the system preserves time-reversal symmetry, the real part of $\sigma^{a,s^i}_{bb}$ corresponds to the spin/charge shift photocurrent.

Our analysis, without loss of generality, will focus on the pure spin photocurrent in a two-dimensional system with time-reversal symmetry, as shown in Fig. \ref{Fig1} (a). Here the zigzag and armchair directions are denoted as the $x$ and $y$ directions respectively. We next examine the symmetry constraints on the conductivity tensor required for the generation of a pure spin photocurrent. In Eq. (\ref{sigma}), the denominator contains terms involving the energy differences $\omega_{mn}({\bf k})$. Under time-reversal symmetry, the energy differences satisfy the relation $\omega_{mn}({\bf k})=\omega_{mn}({-\bf k})$. As a result, the denominators remain invariant in systems that conserve time-reversal symmetry. The numerators, on the other hand, involve terms of the form $j^{a,s^i}_{mn}v^b_{nl}v^b_{lm}$. When the system further preserves the mirror-symmetry along the $x$ direction, as shown in Fig. \ref{Fig1} (a),  the velocity components satisfy the following relationships: $M^x v^x_{mn}(\mathbf{k})=-v^x_{mn}(\mathbf{k^{\prime}})=v^x_{mn}(\mathbf{k})$ and $M^x v^y_{mn}(\mathbf{k})=v^y_{mn}(\mathbf{k^{\prime}})=v^y_{mn}(\mathbf{k})$. Note that $\mathbf{k^{\prime}}$ is also the time-reversal point $-\bf k$ due to the presence of $C_3$ rotational symmetry. Consequently, the charge NLO conductivity tensors $\sigma^{x,s^0}_{xx}$ and $\sigma^{x,s^0}_{yy}$ vanish. The spin current operators satisfy: $M^x j^{x,s^z}_{mn}(\mathbf{k})=j^{x,s^z}_{mn}(\mathbf{k^{\prime}})=j^{x,s^z}_{mn}(\mathbf{k})$ and $M^x j^{y,s^z}_{mn}(\mathbf{k})=-j^{y,s^z}_{mn}(\mathbf{k^{\prime}})=j^{y,s^z}_{mn}(\mathbf{k})$. Therefore, the presence of mirror symmetry $M^x$ can generate a pure spin photocurrent along the $x$-axis direction where the polarization direction of linearly polarized light can be along $x$ or $y$ direction, i.e., $\sigma^{x,s^z}_{xx}\neq 0$ and $\sigma^{x,s^z}_{yy}\neq 0$.

Next, we examine the control of pure spin photocurrent via an applied electric field. An out-of-plane electric field preserves the in-plane mirror symmetry, as shown in Fig. \ref{Fig1} (a), thereby allowing the electrical tuning of the pure spin photocurrent. To achieve this, we begin with the two-band Dirac Hamiltonian, \cite{Golub2011,liu2011Q,liu2011low,Bampoulis2023}
 \begin{equation}
	 	H=\hbar v_f(k_x\hat{\sigma}_x+\hat{\tau}_zk_y\hat{\sigma}_y) + \lambda_{SO}\hat{s}_z\hat{\tau}_z\hat{\sigma}_z+U\hat{\sigma}_z,\label{Ham}
	 \end{equation}
 where $\hbar$ represents Planck's constant and $v_f$ denotes the Fermi velocity. $\hat{\sigma}$ is the Pauli matrix describing the orbital degree of freedom, while $\lambda_{SO}$ represents the intrinsic SOC strength. The operators $\hat{s}_z$ and $\hat{\tau}_z$ correspond to the spin and valley operators, respectively. The parameter $U$ represents the staggered potential, which can be controlled by applying an out-of-plane electric field $E_{\perp}$. After diagonalizing the Hamiltonian in Eq. (\ref{Ham}), the energy spectrum is obtained as follows:
  \begin{equation}
	 	E=\pm \sqrt{(U+\lambda_{SO}s_z \tau_z)^2 +\hbar^2 k^2 v_f^2}.\label{es}
	 \end{equation}
Here, $\tau_z=\pm 1$ and $s_z=\pm1$ label the valley and spin degrees of freedom, respectively. In the regime $\lambda_{SO}>U>0$, the $K$ valley ($\tau_z=1$) hosts spin-down ($s_z=-1$) states at both the valence band maximum and the conduction band minimum, while the $K^{\prime}$ valley ($\tau_z=-1$) exhibits spin-up ($s_z=1$) states at the corresponding band edges. In contrast, when $\lambda_{SO}>-U>0$, a reversal of spin polarization occurs: the $K$ valley now hosts spin-up states, and the $K^{\prime}$ valley hosts spin-down states at both band extrema. 

When the incident light is near the band edge, the excitation and transport of electrons in the vicinity of the $K$ and $K^{\prime}$ valleys are involved. For the regime $\lambda_{SO}>U>0$, the pure spin photocurrent is given by $J^{a,s^z}=\frac{1}{2}(J^{a,s^0}_{K^{\prime}}-J^{a,s^0}_{K})$, where $J^{a,s^0}_{K/K^{\prime}}$ denotes the charge photocurrent generated near the $K/K^{\prime}$ valleys. When the electric field is tuned to the $\lambda_{SO}>-U>0$ regime, a spin reversal occurs in the different valleys, and the pure spin photocurrent becomes $J^{a,s^z}=\frac{1}{2}(J^{a,s^0}_{K}-J^{a,s^0}_{K^{\prime}})$. If the sign of the charge photocurrent in the $K/K^{\prime}$ valley remains unchanged during the transition between the two regimes, the sign of the spin photocurrent can be reversed. Therefore, this spin reversal, induced by switching the sign of the out-of-plane electric field, provides a direct mechanism for electrically controlling the direction of the pure spin photocurrent in systems with mirror symmetry. In the following, we demonstrate this proposed scheme using atomic first-principles calculations for germanene.

\emph{Numerical results} Having established the underlying mechanism for electric-field-tunable pure spin photocurrents, we proceed to validate our theoretical framework based on germanene. To this end, we employ the non-equilibrium Green’s function (NEGF) formalism in conjunction with first-principles calculations based on density functional theory (DFT). Germanene, a group-IV elemental monolayer with a hexagonal lattice analogous to graphene, has been successfully exfoliated from bulk materials in experiments.\cite{Arka2022,li2014b,deng2018,Bampoulis2023} In contrast to the planar structure of graphene, germanene exhibits intrinsic out-of-plane buckling, which provides a pathway for electric field control. Moreover, germanene preserves both inversion symmetry and mirror symmetry along the zigzag direction in its pristine state as shown in Fig. \ref{Fig1}(b). When an out-of-plane electric field is applied, inversion symmetry is broken while mirror symmetry remains intact, enabling the generation of a pure spin photocurrent along the zigzag direction. Next, we systematically investigate the dependence of the pure spin photocurrent on the out-of-plane electric field.



\begin{figure}
	\centering
	\includegraphics[scale=0.45]{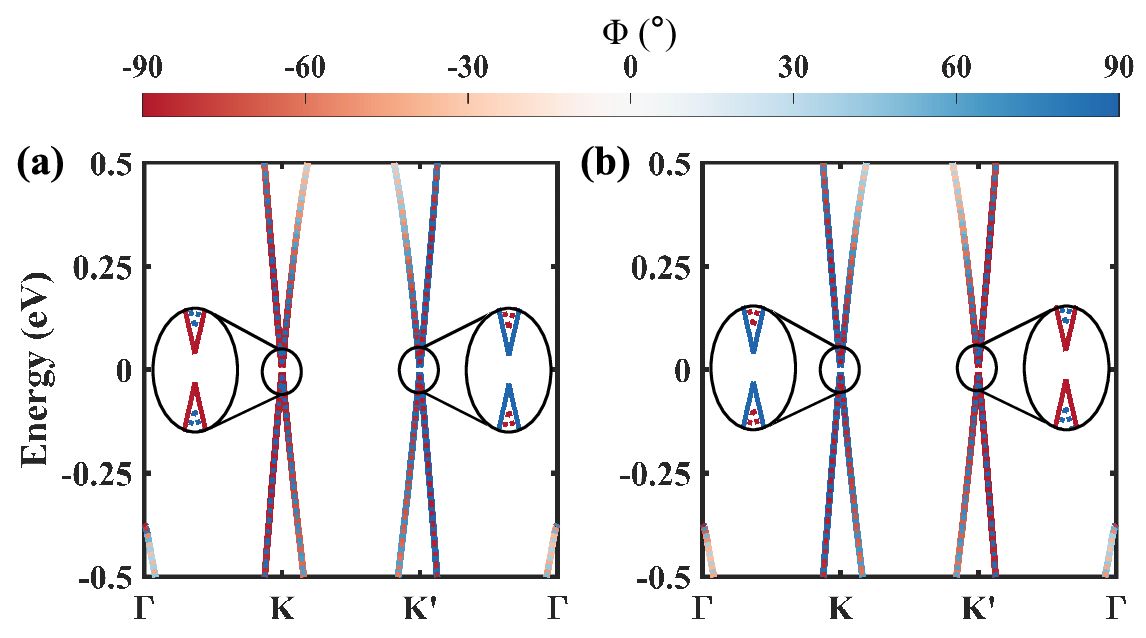}
	\caption{(a, b) Band structure of germanene for gate voltages $V_g=\pm 2$V, respectively. The color scale represents the spin polarization angle $\Phi$, with blue and red indicating spin-up and spin-down states, respectively. }
	\label{Fig2}
\end{figure}

In the absence of a gate voltage, the band structure of germanene exhibits spin degeneracy. However, applying an out-of-plane electric field lifts this degeneracy due to the breaking of inversion symmetry. As shown in Fig. \ref{Fig2}(a), a positive gate voltage results in valence and conduction bands with spin-down characteristics at the $K$ valley, while at the $K^{\prime}$ valley, the valence and conduction bands are spin-up. When a negative gate voltage is applied, these effects are inverted: at the $K$ valley, the valence and conduction bands are spin-up, whereas at the $K^{\prime}$ valley, they are spin-down, as illustrated in Fig. \ref{Fig2}(b). This is consistent with the theoretical analysis in Eq. (\ref{es}). To further elucidate the spin polarization characteristics of germanene, we compute the spin polarization vector ${\bf P}=(P_x,P_y,P_z)$ from the Bloch states and define the spin tilting angle ${\bf P}$ to characterize the spin texture.\cite{Gong2014,zhao2011,yang2021} When $\Phi=\pm 90^\circ$, the spin is fully out-of-plane. Our analysis reveals that under a positive gate voltage, the spin tilting angle $\Phi$ of the valence band maximum approaches $-90^\circ$ near the $K$ point and $90^\circ$ near the $K^{\prime}$ point. Conversely, when a negative gate voltage is applied, the sign of $\Phi$ reverses, in agreement with our previous analysis.

Next, we construct a two-probe transport device based on germanene, as shown in Fig. \ref{Fig1} (b), and analyze the photocurrent generated under linearly polarized light irradiation. In our calculations, the device consists of three regions: a central scattering region and two leads (left and right) that extend to electron reservoirs at infinity. The central scattering region, containing 5.5 unit cells with a total of 45 atoms, spans approximately 22.11 {\AA} in length to maintain mirror symmetry. In the simulation, we add vacuum layers of 9.5 {\AA} along the $z$ direction at the top and bottom of the device to avoid the fake interactions between neighbouring slabs. In the numerical simulation, the $x$ axis runs along the transport direction, and the $z$ axis is perpendicular to the two-dimensional system. Linearly polarized light shines on the central scattering region. In the following, our analysis focuses on the spin-dependent photocurrent flowing into the left lead.  In our calculations, we apply gates across the entire system to tune the pure spin photocurrent generated by linearly polarized light as shown in Fig. \ref{Fig1} (c).


\begin{figure}
	\centering
	\includegraphics[scale=0.51]{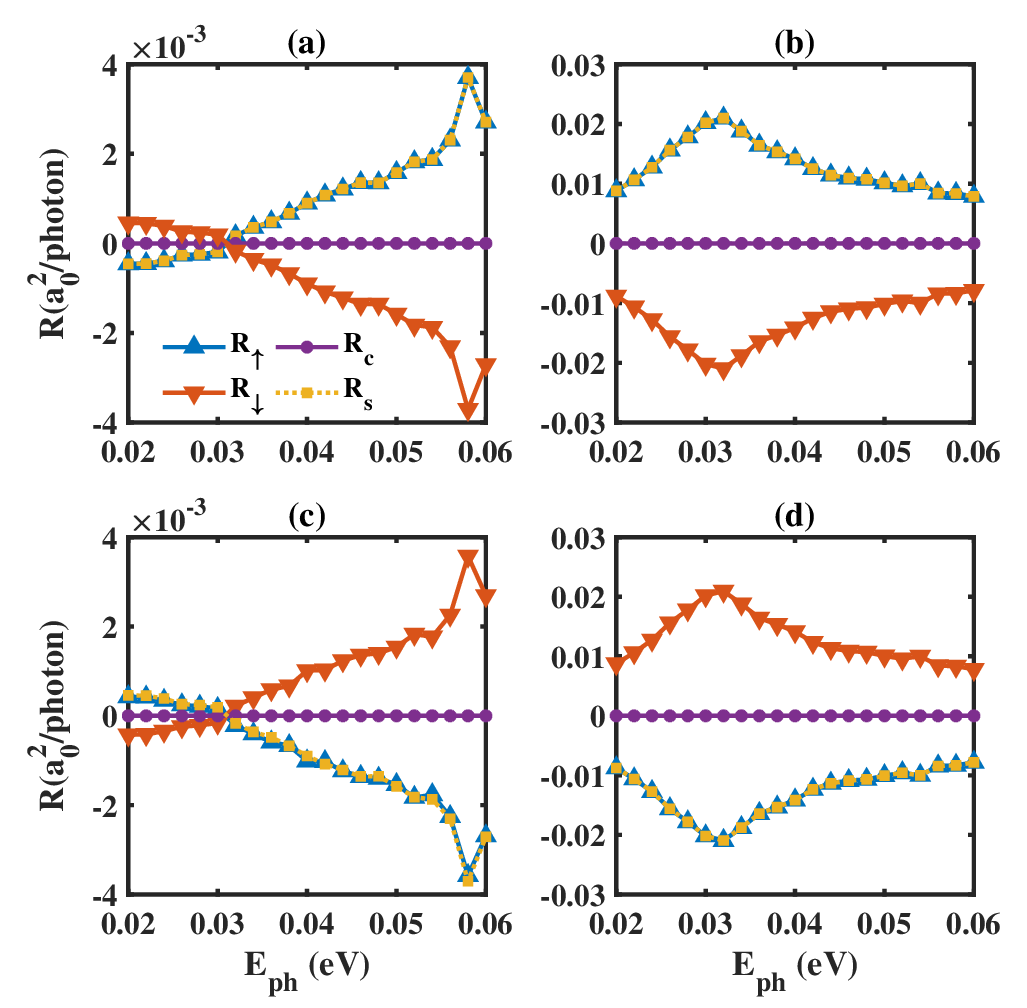}
	\caption{The spin-dependent photoresponse versus the photon energy $E_{ph}$ under vertically incident linearly polarized light irradiation when a $V_g=\pm 2$V electric field is applied. (a, b) The calculated spin-dependent photoresponse under photon polarization along the x-axis and y-axis when applying a gate voltage of $2$ V. (c, d) The calculated spin-dependent photoresponse for photon polarization along the x-axis and y-axis when applying a gate voltage of $-2$ V. 
	}
	\label{Fig3}
\end{figure}

After obtaining the self-consistent device Hamiltonian from NEGF-DFT calculations, the spin-dependent photocurrent $J_{L,s}^{(ph)}$ can be expressed by the following formula,\cite{ivchenko1978,xie2015,chen2018,Chen2012,zhang2014}
\begin{equation}
	\label{photocurrent}
	J^{(ph)}_{L,s}=\frac{ie}{h}\int{{\rm Tr}\{\Gamma_{L}[G^{<}_{ph}+f_L(\epsilon)(G^{>}_{ph}-G^{<}_{ph})]\}_{ss}}d\epsilon,
\end{equation}
where $L$ represents the left electrode and $s$ represents the spin component ($s= \uparrow,\downarrow$); $\Gamma_L=i(\Sigma_L^r-\Sigma_L^a)$ and $f_L(\epsilon)$ denote the linewidth function and the Fermi-Dirac distribution function of the left lead; while $\Sigma^r_L=[\Sigma^a_L]^{\dagger}$ represents the retarded self-energy from the left lead; $G^{</>}_{ph}=G_0^{r}\Sigma_{ph}^{</>}G_0^{a}$ is the lesser/greater Green's function including electron-photon interaction,\cite{henrickson2002} where $\Sigma_{ph}^{</>}$ represents the self-energy from the electron-photon interaction. The light polarization information is incorporated in the self-energy and can be described by a complex vector $\bf{e}$. For linearly polarized light, $\bf{e}=\cos{\theta}\bf{e_1}+\sin{\theta}\bf{e_2}$, where $\theta$ is the angle between the polarization direction and vector $\bf{e_1}$, as shown in Fig. \ref{Fig1} (b). In our numerical calculations, vectors $\bf{e_1}$ and $\bf{e_2}$ are aligned with the $x$ and $-y$ directions respectively, while light is incident along the $-z$ direction at various angles $\beta$, as shown in Fig. \ref{Fig1} (b).

\begin{figure}
	\centering
	\includegraphics[scale=0.51]{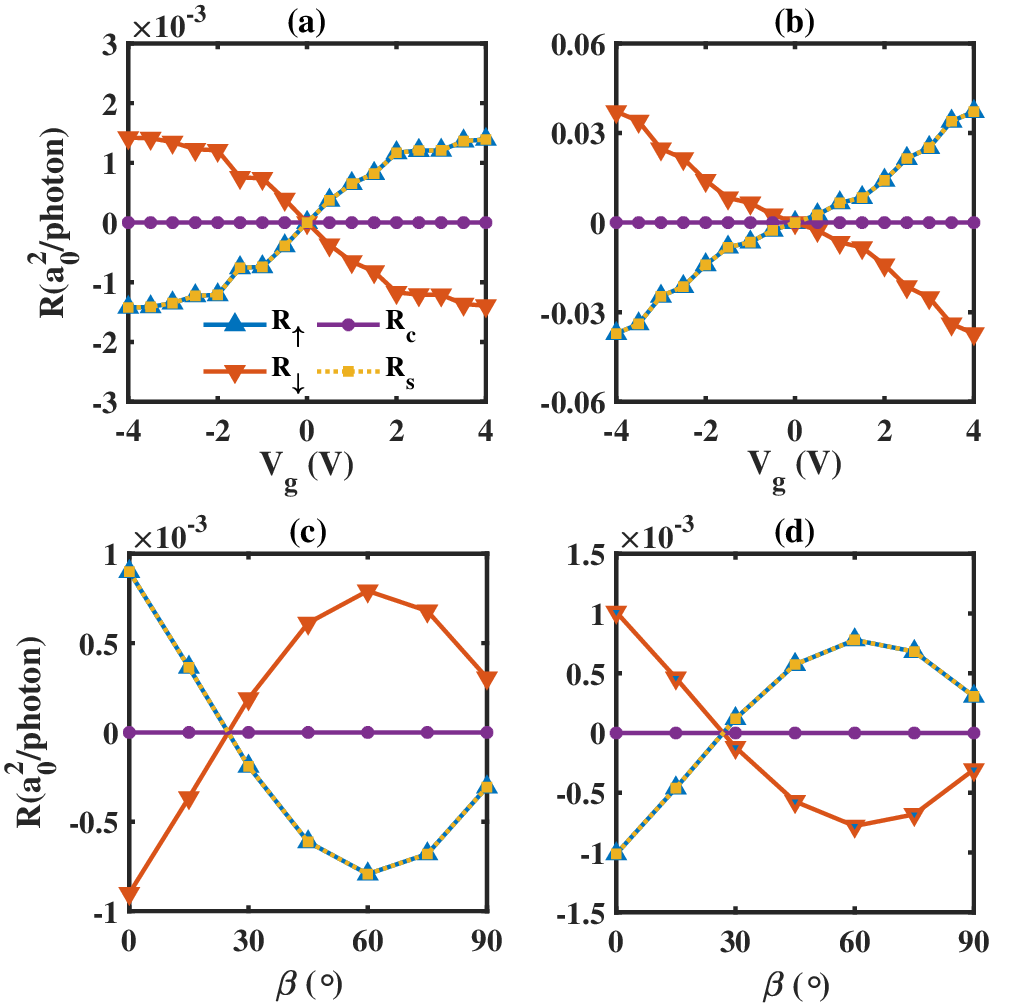}
	\caption{(a, b) The spin-dependent photoresponse versus the gate voltage $V_g$. Here, the incident photon energy is fixed at $E_{ph}$=0.04eV, and the linearly polarized light is polarized along the (a) x-axis and (b) y-axis, respectively. (c, d) The spin-dependent photoresponse versus photon incident angle $\beta$ when applying a gate voltage of $\pm2$ V. Here, the photon energy $E_{ph}$ is equal to 0.04 eV and the linearly polarized light is polarized along the x-axis. 
	}
	\label{Fig4}
\end{figure}

For simplicity, we introduce a normalized photocurrent\cite{henrickson2002,Chen2012,chen2018}, namely the spin-dependent photoresponse $R$,
\begin{equation}
	\label{R_s}
	R_{L,s}=\frac{J^{(ph)}_{L,s}}{eI_\omega},
\end{equation}
where the unit of $R$ is ${a_0^2}/{photon}$ and $a_0$ represents the Bohr radius; $J^{(ph)}_{L,s}$ is the spin-dependent photocurrent defined in Eq. (\ref{photocurrent}); \textit{e} is the electron charge, and $I_\omega$ is the photon flux, defined as the number of photons per unit time per unit area. To better illustrate the photocurrent's spin properties, we also introduced the charge photoresponse $R_{c}=\frac{1}{2}(R_{\uparrow}+R_{\downarrow})$ and spin photoresponse $R_s=\frac{1}{2}(R_{\uparrow}-R_{\downarrow})$.

Next, we'll investigate the photoresponse generated in germanene under linearly polarized light irradiation. Due to the presence of inversion symmetry in germanene in the absence of an external electric field, both charge photocurrent and spin photocurrent are forbidden. To analyze the relationship between photoresponse and incident photon energy, we apply an out-of-plane electric field of $V_g=\pm 2$V as an example. When a voltage of 2V is applied, the spin-up photoresponse gradually increases as the gate voltage rises, while the spin-down photoresponse decreases. Due to mirror symmetry, the magnitudes of the spin-up and spin-down photoresponses are equal but opposite in direction, resulting in zero net charge current. This generates a pure spin photocurrent, as illustrated in Figs. \ref{Fig3} (a, b). Applying a gate voltage in the opposite direction reverses the spin of the generated photoresponse, also shown in Figs. \ref{Fig3} (c, d). This indicates that when an external electric field is applied, pure spin photocurrent can always be generated regardless of the photon energy of the incident polarized light. These numerical results are in agreement with the theoretical predictions derived from the NLO response analysis. Since the spin direction of the generated pure spin photocurrent is opposite when gate voltages are applied in different directions, we will analyze how gate voltage affects the spin photocurrent.

Next, we examine the relationship between photocurrent and external electric field by calculating the photocurrent variation with gate voltage under linearly polarized light along the $x$- and $y$-axes. As shown in Figs. \ref{Fig4} (a, b), it can be observed that mirror symmetry forbids charge current while allowing non-zero spin photocurrent, resulting in spin-up and spin-down spin-dependent photoresponses having equal magnitude but opposite directions. As the applied electric field increases, the magnitude of the generated spin photocurrent gradually rises. Furthermore, when the gate voltage direction reverses, the spin photocurrent also reverses direction. Finally, we also investigate the effect of the incident angle of linearly polarized light on spin-dependent photoresponse, as shown in Figs. \ref{Fig4} (c, d). It was found that by adjusting the incident angle of the incident light, the photocurrent could be adjusted while still maintaining its pure spin characteristic.

\emph{Summary}-In summary, we propose a theoretical scheme to control pure spin photocurrent using an electric field in time-reversal symmetric systems with mirror symmetry. Using first-principles calculations, we verified the feasibility of this scheme in germanene, a material that has been experimentally realized. By adjusting the external electric field, we can tune both the magnitude and direction of pure spin photocurrent along the zigzag direction when the light polarization is along either $x$ or $y$ direction. These findings offer new possibilities for low-power, volatile spin-based memory and optospintronic devices.

$${\bf ACKNOWLEDGMENTS}$$

We gratefully acknowledge the support from the National Natural Science Foundation of China (Grant No. 12474047, 12174231, 12404057, 12404151), the Fund for Shanxi ``1331 Project", Research Project Supported by Shanxi Scholarship Council of China. This research was partially conducted using the High Performance Computer of Shanxi University.

\section{Appendix: Computational details of DFT}
Before analyzing the spin-dependent photocurrent, the atomic structures of germanene are fully relaxed using the Vienna ab initio simulation package (VASP)\cite{Kresse1993,Kresse1996} until the forces acting on each atom are less than 0.01 eV{\AA}$^{-1}$. The first Brillouin zone of the unit cell is sampled by a 15$\times$15$\times$1 $\bf k$ space grid during the calculation, with an energy cutoff of 600 eV used for the plane-wave expansion to ensure accuracy. The optimized lattice constant of germanene is 4.673 {\AA}. The Hamiltonian calculation was performed using the Keldysh non-equilibrium Green's function (NEGF) formalism\cite{taylor2001,zhang2014} combined with density functional theory (DFT), implemented in the first-principles quantum transport package Nanodcal\cite{nanodcal2,zhang2014,taylor2001}. In the NEGF-DFT self-consistent calculation, we use the linear combination of atomic orbital (LCAO) basis at the double-$\zeta$ polarization (DZP) level to expand the wave functions and other physical quantities, while the atomic cores are described by standard norm-conserving non-local pseudo-potentials\cite{kleinman1982,perdew1996}. The NEGF-DFT self-consistency is achieved when the difference between iterative steps of the monitored quantities (including all elements of the Hamiltonian and density matrices) becomes less than 1$\times$10$^{-8}$ a.u. The first-principles quantum transport calculations are carried out in two steps: first, the Hamiltonian of the two-probe device without electron-photon interaction is obtained through NEGF-DFT self-consistently\cite{taylor2001}; second, the electron-photon interaction is perturbatively included as self-energy in terms of NEGF during photocurrent calculation\cite{zhang2014}.

\bigskip

\noindent{$^{*)}$chenjun@sxu.edu.cn}\\
\noindent{$^{\dagger)}$zhanglei@sxu.edu.cn}

\bibliography{ref}

\end{document}